\documentclass[sigconf]{acmart}
\usepackage{graphicx}               
\usepackage{todonotes}
\usepackage[binary-units]{siunitx}
\usepackage{multirow}

\usepackage{soul}
\AtBeginDocument{%
  }

\acmConference[?]{?}{?}{?}

\begin{document}

\title{Reducing Memory Requirements for the IPU \\ using Butterfly Factorizations}


\author{S.-Kazem Shekofteh}
\affiliation{%
  \institution{Computing Systems Group}
  \city{Heidelberg}
  \country{Germany}}
\email{kazem.shekofteh@ziti.uni-heidelberg.de}

\author{Christian Alles}
\affiliation{%
  \institution{Computing Systems Group}
\city{Heidelberg}
\country{Germany}}
\email{christian.alles@stud.uni-heidelberg.de}

\author{Holger Fröning}
\affiliation{%
	\institution{Computing Systems Group}
	\city{Heidelberg}
	\country{Germany}}
\email{holger.froening@ziti.uni-heidelberg.de}
 
\renewcommand{\shortauthors}{Shekofteh et al.}

\begin{abstract}

High Performance Computing (HPC) benefits from different improvements during last decades, specially in terms of hardware platforms to provide more processing power while maintaining the power consumption at a reasonable level. 
The Intelligence Processing Unit (IPU) is a new type of massively parallel processor, designed to speedup parallel computations with huge number of processing cores and on-chip memory components connected with high-speed fabrics. 
IPUs mainly target machine learning applications, however, due to the architectural differences between GPUs and IPUs, especially significantly less memory capacity on an IPU, methods for reducing model size by sparsification have to be considered. Butterfly factorizations are well-known replacements for fully-connected and convolutional layers.
In this paper, we examine how butterfly structures can be implemented on an IPU and studiy their behavior and performance compared to a GPU.
Experimental results indicate that these methods can provide 98.5\% compression ratio to decrease the immense need for memory, the IPU implementation can benefit from 1.3x and 1.6x performance improvement for butterfly and pixelated butterfly, respoectively. We also reach to 1.62x training time speedup on a real-word dataset such as CIFAR10.

\end{abstract}

\ccsdesc[500]{Computing methodologies~Massively parallel algorithms}
\ccsdesc[300]{Theory of computation~Parallel computing models}
\ccsdesc[300]{Computer systems organization~Multiple instruction, multiple data}
\keywords{Massively Parallel Processing, Sparse Data Structure, Intelligence Processing Units}

\received{09 August 2023}
\received[revised]{09 August 2023}
\received[accepted]{09 August 2023}

\maketitle

\section{Introduction}
In recent years, 
accelerators such as GPUs have become widely adopted across various scientific disciplines as a means of improving performance, as they can offer tremendous performance improvements for certain tasks. 
However, such a specialization comes at reduced general applicability, thus the future is anticipated as massively heterogeneous.
In this regard, the Intelligence Processing Unit (IPU) represents a novel form of accelerator, particularly designed to meet the demands of Machine Learning (ML) applications based on deep neural networks. 
Unlike the single instruction multiple threads (SIMT) architecture commonly found in GPUs, which relies on contiguous vectorized data for optimal performance, the IPU explicitly targets both dense and irregular sparse data access. 
This is achieved by executing individual processing threads on small data blocks and leveraging a multiple instruction multiple data (MIMD) architecture \cite{ipu1}.

Most literature dealing with the IPU uses this processor to accelerate ML applications with frameworks such as TensorFlow and PyTorch. A common insight is the limited amount of memory and the high computational performance potential. Specifically, \cite{ipu_ML_bench1}, \cite{Dyn_Sparse}, and \cite{Qwant} try to exploit the IPUs for running and evaluating machine learning algorithms, so that they can benefit from the high computational performance of the IPU. More precisely, \cite{ipu_ML_bench1} talks about key differences between GPUs and IPUs when running ML algorithms for neural network training and inference.

Machine learning applications involve various transformation steps, such as the discrete Fourier transform (DFT) and discrete cosine transform (DCT), which are primarily based on the fast Fourier transform (FFT) algorithm. 
To eliminate the need for manually crafting implementations of these transformations, one can propose an approach where the most suitable transform for a specific task and dataset is learned automatically. 
In this regards, a particular factorization using sequences of special block diagonal matrices, called butterfly factorization \cite{butterfly}, was proposed that results in a class of structured matrices with $O(N)$ parameters and automatic fast multiplication in $O(N \log N)$ operations as a replacement for the $O(N^2)$ algorithm of dense MM.


Overall, this work is concerned with the analysis of butterfly factorization for the execution of neural networks on an IPU.
First, we assess the basic performance characteristic of the IPU, in contrast with a GPU comparable in terms of performance, power, and year of introduction.
We find that memory capacity is indeed a scarce resource on an IPU, and highlight some situations where additional constraints occur.
These findings motivate the need for methods to reduce memory footprint on an IPU, for which we propose to use Butterfly factorizations.
Smaller memory footprints would allow to execute larger neural architectures. 
However, factorizations come with overhead in computation, thus it is important to assess the trading among reduced memory requirements and increase in computational overhead.
Only with a sound compromise promising solutions can be found for overall performance. In detail, this work makes the following contributions:
\begin{enumerate}
	\item We provide a detailed analysis of the IPU performance for linear algebra operations, including dense and sparse matrix multiply as well as operations on skewed matrices. We highlight some situations where unexpected additional demand on memory allocations happens.
	\item We assess the performance of two butterfly factorization variants on the IPU in terms of memory footprint reduction and computational overhead, showing that GPU optimizations for butterfly factorization are not necessarily helpful for other accelerators such as the IPU.
	We provide a brief discussion of an analysis of parameter effects and possible optimizations for butterfly on the IPU.
\end{enumerate}



Section two delves deeper into background and related work, while section three elucidates the research motivation. Section four outlines implementations and benchmark results, and section five analyzes the impact of tunable parameters on IPU performance metrics. Lastly, section six presents the conclusion.
\section{Background and Related Work}

\subsection{Comparsion of IPU and GPU}

The Graphcore Intelligence Processing Unit (IPU) is specifically engineered to excel in highly parallel processing tasks, making it exceptionally well-suited for handling both dense and sparse workloads across a broad spectrum of applications. This includes areas such as machine learning and graph processing algorithms, where the inherent parallelism of IPU in combination with support for sparsity is  advantageous.

In this work, an M2000 IPU Pod-4 system containing four GC200 IPU processors is used. 
To find a suitable GPU for the comparison, we searched for a similar release year, power envelope and technology. 
The GC200 IPU has been released in July 2020 with a TDP of 150W and a transistor size of 7nm. 
The chosen counterpart GPU is a NVIDIA A30, which was released in April 2021 with a TDP of 165W and with the same transistor technology of 7nm.
Table \ref{table:GC200vsA30} provides a comparision of the specs.

\begin{table}[htbp]
    \caption{Comparison of Graphcore GC200 and NVIDIA A30}
    \label{table:GC200vsA30}
    \centering
	\begin{tabular}{l c c} 
		\hline
		 & A30 & GC200  \\  
		\hline
		Number of cores 		 	& 3584 & 1472 \\
		On-chip memory 			 	& 10.75 MB & 900 MB \\
		On-chip memory bandwidth 	& 5.5 TB/s & 47.5 TB/s \\
		Off-chip memory 			& 24 GB & 64 GB\\
		Off-chip memory bandwidth 	& 933 GB/s & 20 GB/s\\ 
		Inter-Chip bandwidth 		& 200 GB/s & 320 GB/s \\
		FP32 peak compute 			& 10.3 TFLOPS & 62.5 TFLOPS \\
		TF32 peak compute 			& 82 TFLOPS & - \\
		Clock frequency 			& 1.44 GHz & 1.33 GHz \\ 
		Power Consumption 			& 165 W & 150 W  \\
		
		\hline
	\end{tabular}
	 \vspace{-4mm}
\end{table}

The Graphcore IPU consists of two basic building blocks: IPU-Tiles and the IPU-Exchange. 
Each IPU-Tile consists of an IPU-Core and In-Processor-Memory. In contrast to a GPU executing in a SIMD fashion, each IPU-Core can schedule six threads in a time-sliced fashion. 
Therefore, it is capable of executing in an MIMD fashion. The In-Processor-Memory of an IPU contains a small amount of SRAM memory, and as many such units exist, a large amount of SRAM in total is present.

Graphcore supports ML frameworks such as Tensorflow and PyTorch, along with its Poplar C++ framework for low-level programming. 
IPU-Programs are represented as dataflow graphs, with computation as nodes (Vertices) and data as Tensors connected via edges. 
Vertices can be mapped to tiles and executed independently. The graph can be created via Poplar API or Poplibs libraries. 
Poplar requires explicit mapping, while PopLibs provide functions for common operations and manage vertex mapping and data copying.

\subsection{Related work on the IPU}

Earlier work in the context of the IPU mostly deals with accelerating various ML workloads. 
Kacher et al. \cite{kacher2020graphcore} evaluate the performance of the first generation GC2 IPU for deep neural networks. 
Sumeet et al. \cite{sumeet2022performance} analyze the performance of an M2000 consisting of four second generation GC200 IPUs for text detection applications.
Vaswani et al. \cite{vaswani2022confidential} present Trusted Extensions, a set of experimental hardware extensions to allow running AI workloads with strong confidentiality and integrity guarantees. 

Going beyond ML, Luow and McIntosh-Smith \cite{louw2021using} use the first-generation GC2 IPU for stencil computations on structured grids and characterize the performance with STREAM memory benchmark results and a roofline model. 
They inspect Poplar and reported sufficient programmability to implement these HPC problems, and achieve performance compared to that of modern GPUs.

Jia et al. \cite{jia2019dissecting} dissect the IPU with different microbenchmarks. 
They examine a GC2 IPU-Pod16 system 
in terms of computational performance, peak bandwidth (inter and intra IPU) and other metrics. 
They also reported that, in single precision, a single IPU processor outperforms a V100 by factor two (31.1 TFlops for the IPU and 15.7 TFlops for the GPU). 

The study by Balewski et al. \cite{M2000vaA100PMBS2022} compares time-series ML-based regression on IPUs against GPUs. 
They conduct a comparison of training performance between an M2000 system and an NVIDIA A100 GPU system, including multiple convolutional and fully-connected models with batch normalization.
Their results show a very similar performance on both devices. 
Although the achieved validation loss on the IPU was only 10\% to 20\% larger, IPU provides 2.5 to 3 times smaller energy consumption per epoch than the GPU.

\subsection{Butterfly factorizations}

In the field of ML, fast linear transform functions such as the DFT or DCT are used to speed up training and inference. 
The underlying problem is to find an appropriate transform with an efficient implementation for a given model. 
Realizing such an algorithm for a given framework and platform is often tedious. 
As every structured linear transform, including convoluational and fully-connected layers, can be described with dense matrix-vector multiplication, Dao et al.~\cite{butterfly} propose butterfly matrices, replacing specific transformations by universal building blocks called butterfly factors.

\subsubsection{Butterfly}

These $O(\log N)$ butterfly factors, each consisting of $O(N)$ nonzero entries, are being multiplied, resulting in an $O(N \log N)$ algorithm as a replacement for the $O(N^2)$ algorithm of dense MM.
As a result, both the execution time (by having a more efficient algorithm) and the memory footprint (by applying sparsification) are reduced. 
By making the nonzero entries of the butterfly factors learnable, a variety of linear transforms can be represented by the butterfly factorization.

\begin{figure}
	\centering
	\includegraphics[width=1\linewidth]{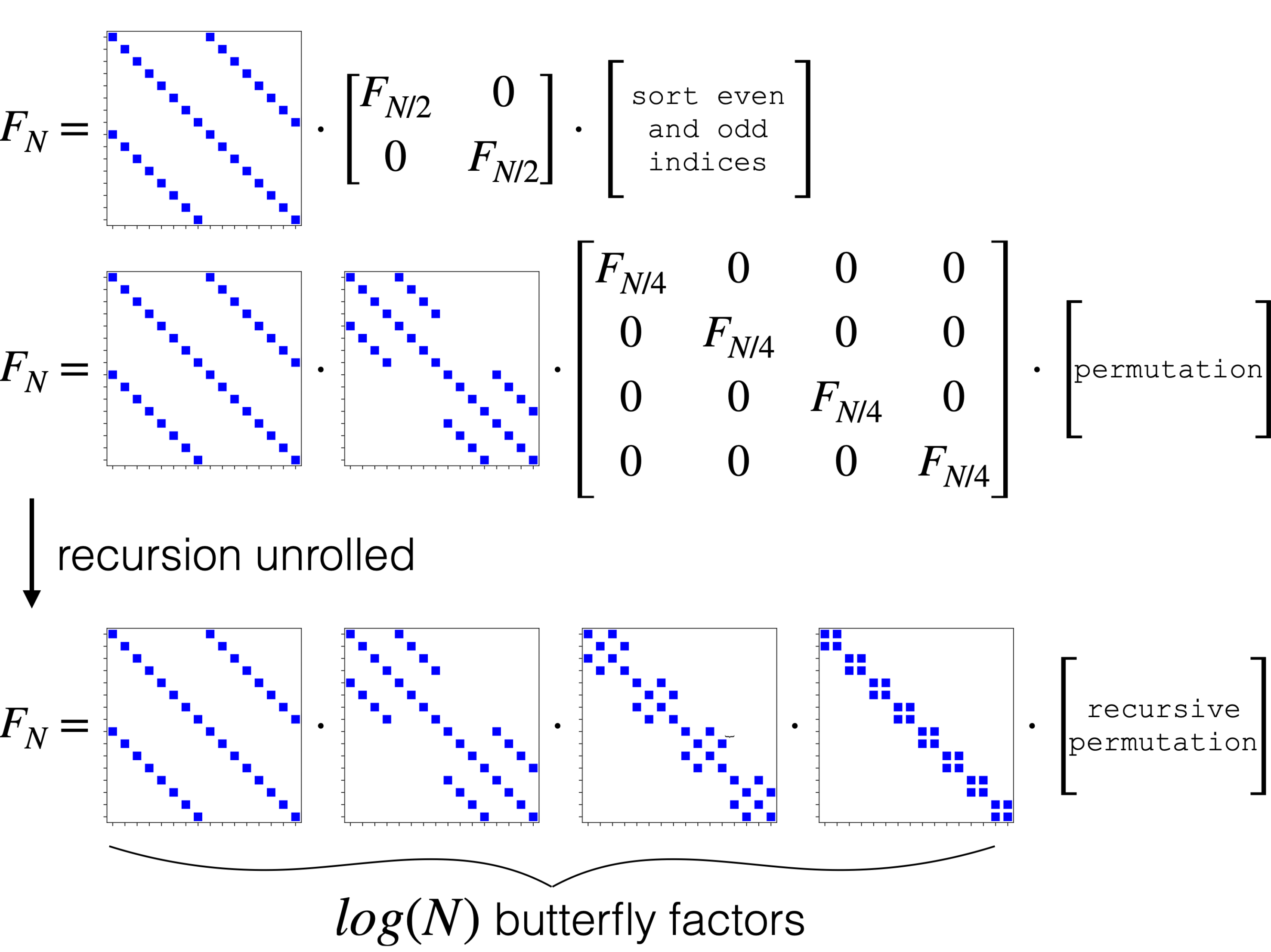}
	\caption{Expressing an input matrix as the product of $\log N$ butterfly factors. 
		This FFT-representation also applies to the compression of structured matrices in deep neural networks.
	}
	\label{fig:butterfly}
\end{figure}

Figure \ref{fig:butterfly} shows an example of butterfly factorization. 
Butterfly factorization is inspired by the Cooley-Tukey FFT algorithm. 
Both approaches apply the same divide-and-conquer strategy to generate the butterfly factors. 
Equation \ref{eq:FFT} shows the Cooley-Tukey FFT algorithm for Fig. \ref{fig:butterfly} with matrix representation, while equation \ref{eq:Butterfly} shows the butterfly factorization with matrix representation.

\begin{equation}
	\label{eq:FFT}
	F_{N}=\left[\begin{array}{c c}{{I_{N/2}}}&{{\Omega_{N/2}}}\\ {{I_{N/2}}}&{{-\Omega_{N/2}}}\end{array}\right]
	\left[\begin{array}{c c}{{F_{N/2}}}&{{0}}\\ {{0}}&{{F_{N/2}}}\end{array}\right]
	\left[\begin{array}{c c}
	\parbox[t]{1.6cm}{\centering\small{sort even and odd indices}}
	\end{array}\right]
\end{equation}

\begin{equation}
	\label{eq:Butterfly}
	T_{N}=\left[\begin{array}{c c}{{D_1}}&{{{D_2}}}\\ {{D_3}}&{{D_4}}\end{array}\right]
	\left[\begin{array}{c c}{{T_{N/2}}}&{{0}}\\ {{0}}&{{T_{N/2}}}\end{array}\right]
	\left[\begin{array}{c c}
	\parbox[t]{2.1cm}{\centering\small{separated into two halves by some permutation}}
	\end{array}
	\right]
\end{equation}

Therefore, equation \ref{eq:FFT} is a special case of \ref{eq:Butterfly} where $D_1$ = $D_3$ = $I_{N/2}$, $D_2$ = $\Omega_{N/2}$, $D_4$ = $-\Omega_{N/2}$, $F_{N/2}$ = $T_{N/2}$ and the permutation is the separation of even and odd indices. 
This translates to every structured matrix being able to be decomposed into $\log N$ butterfly factors in which the sparsity comes from the FFT factorization as shown in Fig.\ref{fig:butterfly}.
Last, unrolling \ref{eq:Butterfly} results in the butterfly factorization:
\begin{equation}
	\label{eq:Butterfly_unrolled}
	T_{N} = B^{(N)}P^{(N)}
\end{equation} 
with $B^{(N)}$ being a butterfly matrix and $P^{(N)}$ being a permutation.

\subsubsection{Pixelated Butterfly (Pixelfly)}

The Pixelated Butterfly approach from \cite{chen2021pixelated} is based on the butterfly matrices with additional efficiency improvements, namely Flat Block butterfly, and additional low-rank terms. 
Flat Block butterfly consists of two additions to the butterfly factorization:
\begin{enumerate}
	\item Flat butterfly approximates the butterfly factorization by replacing the product by a sum with residual connections, enabling easier parallelization. 
	\item Block butterfly takes into account the block data access of a GPU by aligning the butterfly factors, thereby reducing memory accesses. 
\end{enumerate}

\begin{figure*}[!h]
	\centering
	\includegraphics[width=.7\textwidth]{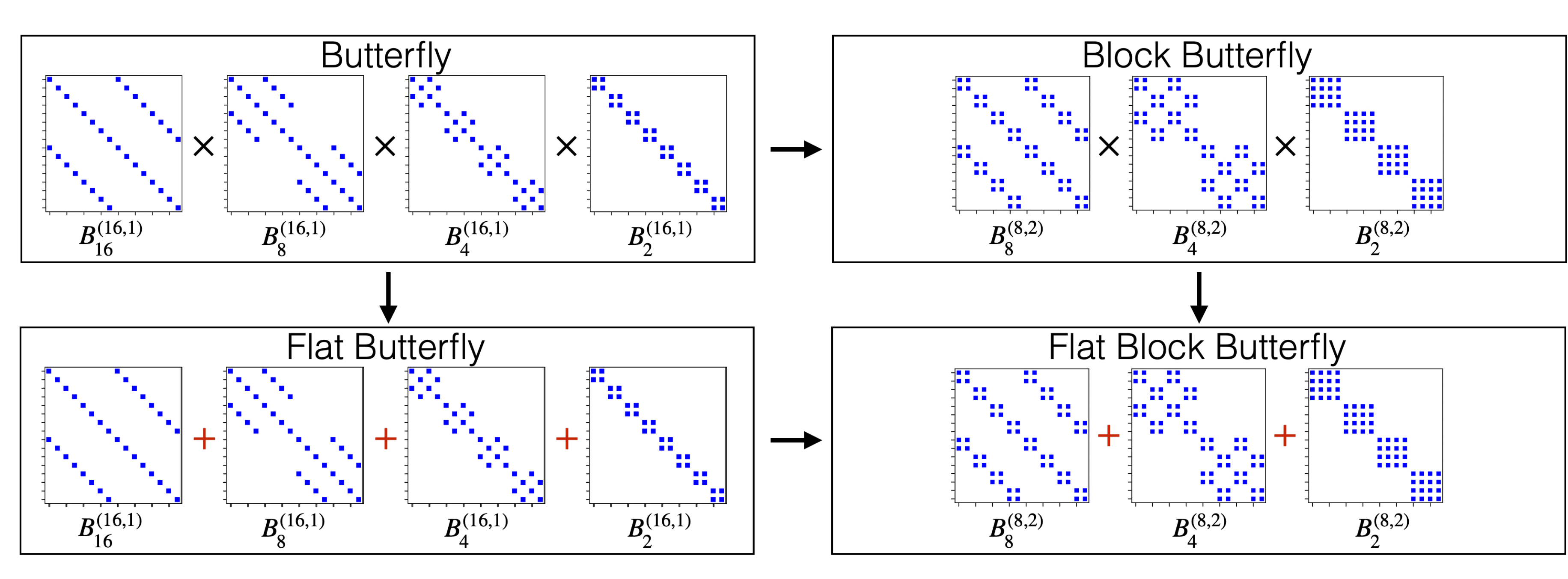}
	\caption{Flat Block butterfly construction by block-aligning the butterfly factors (block butterfly) and approximating the products of butterfly factors by sums (flat butterfly). 
	}
\label{fig:pixelfly}
\end{figure*}

Figure \ref{fig:pixelfly} visualizes Flat Block butterfly with its components. In contrast to the butterfly implementation, the implementation of \cite{chen2021pixelated} for pixelfly relies on either Triton or HuggingFace as a backend. 
However, we restrict ourselves to plain PyTorch due to the absence of some frameworks for the IPU. 
To overcome this issue, we revert to using the implementation from \cite{PixelflyTorch}. 
In contrast to the butterfly factorization, which, like the fully-connected layer, needs only the dimensions as input parameters, the Pixelfly adds additional parameters for configuration. 
These are the size for the low-rank decomposition, the block size and the butterfly size. As we will see later, this directly influences memory footprint and execution time.


We extend the findings of Dao et al. \cite{butterfly} by conducting experiments with butterfly factorization on the second generation GC200 IPU and the A30 GPU, with the Tensor Cores turned off and on. 
Furthermore, we add additional scenarios, layers and performance metrics to the experiments.
Pixelated butterfly as presented by Chen et al. \cite{chen2021pixelated} incorporates properties to adapt butterfly factorization to the GPU architecture. 
They show that the performance of butterfly can be increased with Flat Block butterfly and low-rank terms. 
Similarly, we extend their findings by analyzing the performance and memory size reduction by conducting experiments on the GC200 IPU and A30 GPU.

In summary, this work differs from previous work by analyzing the performance of the second-generation IPU (GC200), while most previous work was concerned with the first generation (GC2).
In this regard, a prime question at hand is to which extend previous findings hold true for the current generation.
Futhermore, we analyze butterfly factorizations on a sparse accelerator such as the IPU, while previous work focused on dense accelerators such as a GPU.


\section{IPU GC200 Performance Analysis}

In the following, we discuss the potentials of using IPU and find appropriate applications and cases where the IPU might be beneficial compared to a GPU. In this regard, we first look at the IPU-Exchange, the ultra-fast and jitter-free communication technology to see how it handles data movements. Then, we inspect the performance of matrix multiplication (MM) operations, as the core of most of parallel applications including deep neural networks. Finally, we study the limitations of memory and its effects on the IPU performance.

The M2000 IPU-Machine used in this work consists of four GC200 IPUs, the second generation IPU, improving overall memory capacity and the number of cores compared to the first generation GC2 IPU processor \cite{graphcoreProcessors}. As this type of hardware is quite novel with the first generation being released in 2018 \cite{GC2Release} and to make the comparison to the GPU fairer, we restrict ourselves to a single IPU.

\subsection{Locality in communication (IPU-Exchange)}

\begin{figure}[htbp]
	\centerline{\includegraphics[width=\columnwidth]{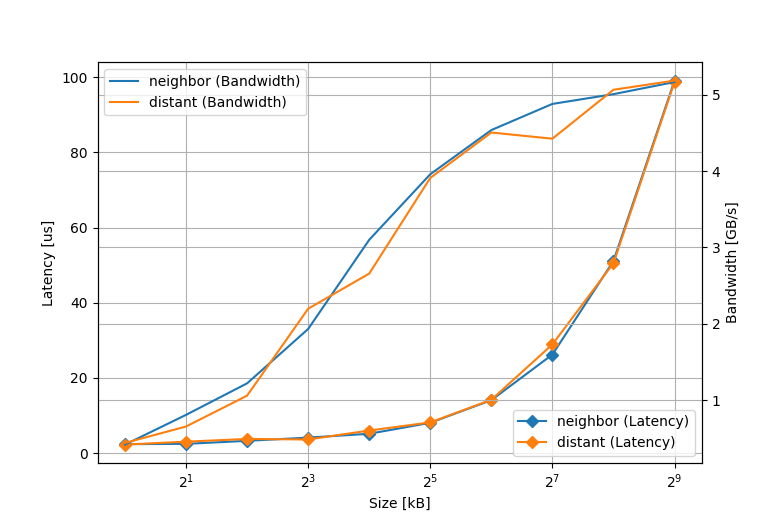}}
	\caption{Latency and Bandwidth within a GC200 IPU with different physical proximity. The pair of neighboring IPU-Tiles is (0,1) and the chosen pair of distant IPU-Tiles is (0,644)}
	\label{fig: IPU_BW}
\end{figure}

IPU and GPU have distinct memory architectures, leading to varying thread memory capacities. A GPU thread accesses GBs of data, whereas a GC200 IPU thread can only access \qty{611}{\kilo\byte}. Hence, data exchanges between IPU-Tiles are more evident than within the GPU. To examine how data movement affects the performance of the IPU, we analyze the memory bandwidth and memory access costs to figure out the effect of data locality, in terms of distance from IPU-Tiles. In this experiment, we evaluate if accessing a data element from either a neighboring IPU-Tile or a far distant one will have different performance or not. The results are depicted in Fig. \ref{fig: IPU_BW}, where it confirms that in case the data does not fit in the Tile-Memory of a given IPU-Tile, it does not matter where the data is stored as long as it fits onto the entire processing unit.

\textbf{Observation 1}: Latency and bandwidth of data accesses in between different IPU-Tiles are tightly coupled with data size, but are independent of their location.

\subsection{Dense and sparse linear algebra}

In order to compare the performance of A30 GPU and GC200 IPU, we implement and evaluate different variants of matrix multiplication (MM) including dense $\times$ dense, sparse $\times$ dense, square, and skewed MM. 
First, we are concerned with square matrices, and compare dense and sparse MM for such.


\begin{table*}[]
	\caption{Performance evaluation of dense vs sparse matrices on GPU vs IPU with different configurations and tools: numbers in GFlops ( \{GPU Peak: FP32: 10300, TF32: 82000\} ,  \{IPU Peak: 62500\} )}
	\label{tab:GPUvsIPU}
	\resizebox{\textwidth}{!}{%
		\begin{tabular}{cccccccccc|cccc}
			\hline
			\multicolumn{10}{c|}{Dense Matrices}                                                                                                                                                                                                                                                                                                                                                                                                                                                                                                                                                                                                                                                                                                                                                          & \multicolumn{4}{c}{Sparse Matrices}                                                                                                                   \\ \hline
			\multicolumn{4}{c|}{GPU low-level}                                                                                                                                                                                                                                                                                            & \multicolumn{3}{c|}{IPU low-level}                                                                                                                                                                                                & \multicolumn{2}{c|}{GPU high-level}                                                                                                                           & \begin{tabular}[c]{@{}c@{}}IPU \\ high-level\end{tabular} & \multicolumn{2}{c|}{GPU cusparse}                                                    & \multicolumn{2}{c}{IPU popsparse}                              \\ \hline
			\multicolumn{1}{c|}{\begin{tabular}[c]{@{}c@{}}GPU\\ naive\end{tabular}} & \multicolumn{1}{c|}{\begin{tabular}[c]{@{}c@{}}GPU\\ shmem\end{tabular}} & \multicolumn{1}{c|}{\begin{tabular}[c]{@{}c@{}}GPU\\ cublas\\ (FP32)\end{tabular}} & \multicolumn{1}{c|}{\begin{tabular}[c]{@{}c@{}}GPU\\ cublas\\ (TF32)\end{tabular}} & \multicolumn{1}{c|}{\begin{tabular}[c]{@{}c@{}}IPU\\ naive\end{tabular}} & \multicolumn{1}{c|}{\begin{tabular}[c]{@{}c@{}}IPU\\ blocked\end{tabular}} & \multicolumn{1}{c|}{\begin{tabular}[c]{@{}c@{}}IPU\\ poplin\end{tabular}} & \multicolumn{1}{c|}{\begin{tabular}[c]{@{}c@{}}PyTorch\\ (FP32)\end{tabular}} & \multicolumn{1}{c|}{\begin{tabular}[c]{@{}c@{}}PyTorch\\ (TF32)\end{tabular}} & PopTorch                                                  & \multicolumn{1}{c|}{99\%}           & \multicolumn{1}{c|}{90\%}                       & \multicolumn{1}{c|}{99\%}            & 90\%                      \\ \hline
			\multicolumn{1}{c|}{1091}                                                & \multicolumn{1}{c|}{2076}                                                & \multicolumn{1}{c|}{9722}                                                          & \multicolumn{1}{c|}{59312}                                                & \multicolumn{1}{c|}{525}                                                 & \multicolumn{1}{c|}{93}                                                    & \multicolumn{1}{c|}{44219}                                                & \multicolumn{1}{c|}{9286}                                 & \multicolumn{1}{c|}{58146}                                 & {1677}                                  & \multicolumn{1}{c|}{\textbf{93215}} & \multicolumn{1}{c|}{10817*} & \multicolumn{1}{c|}{\textbf{76231*}} & {22845} \\ \hline
		\end{tabular}%
	}
 
	\begin{minipage}{\textwidth}

		\small  Note 1: In each column, the best performance out of a set of experiments is show and s bold number represents surpassing the device's peak.
		\\
		\small  Note 2: For sparse matrices, there exist two different implementations: CSR and COO. On both GPU and IPU, CSR shows better performance.
		\\
		\small  Note 3: Performance of IPU blocked suffers from too much temporal data being allocated and many copies taking place.
		\\
		\small  Note 4: PopTorch performance numbers inherertly include data copy time, because PopTorch does not allow to separate the graph in contrast to poplar.
		
	\end{minipage}
\end{table*}

The results for dense and sparse matrices are summarized in Table \ref{tab:GPUvsIPU}. 
As one can see, the IPU surpasses the GPU as long as the matrices can fit into the IPU memory and Tensor Cores (TC) are turned off on the GPU. 
This confirms previous expectations, but highlight the benefits (high computational performance) 
of the IPU, in particular in comparison to a similar accelerator such as the A30.


We also studied the effect of skewness on both GPU and IPU. 
Formally, for a matrix multiplication represented by $A^{(m\times n)} \times B^{(n\times k)} = C^{(m\times k)}$, the skewness ratio for matrix $A$ is defined as $s=\frac{m}{n}$.
Many operations found in deep neural networks are depending in their dimensions on the workload at hand, for instance mini-batch size, size of a fully-connected layer but also the configuration of a convoluational layer in terms of filter size, number of filters (equals the number of output channels) and number of input channels.

\begin{figure*}
	\centerline{\includegraphics[width=0.85\textwidth]{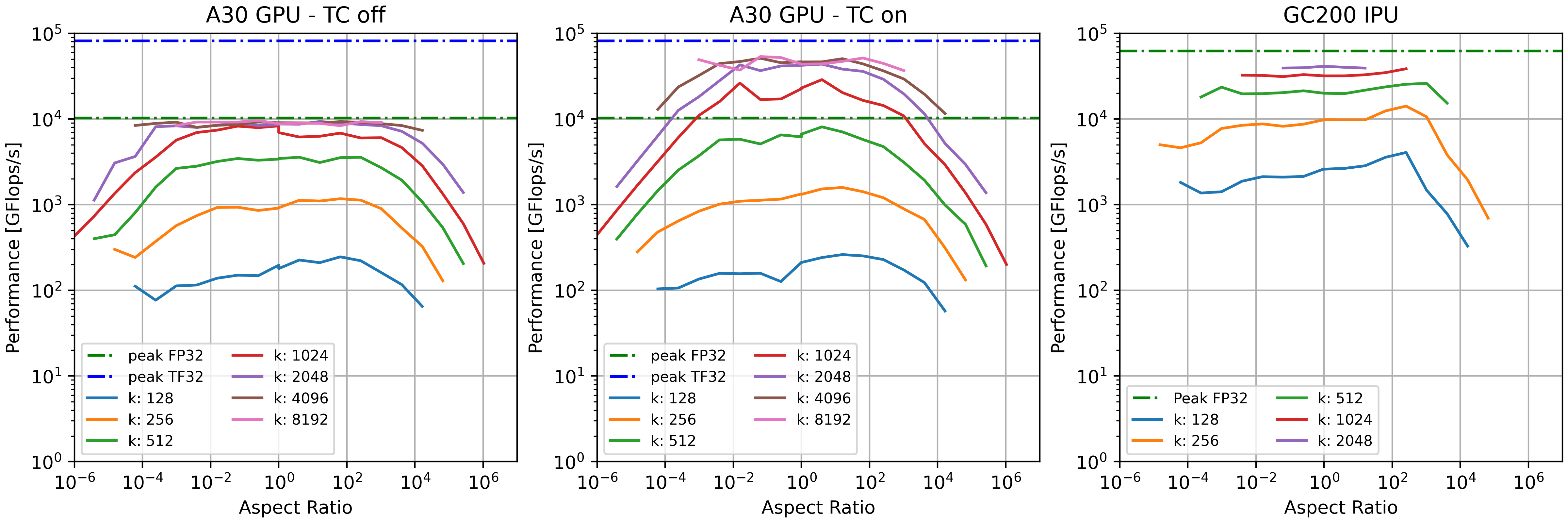}}
	\caption{Skewed MM on GPU vs. IPU}
	\label{fig: skewed_mm_gpu_ipu}
\end{figure*}

Fig. \ref{fig: skewed_mm_gpu_ipu} shows results on skewed matrix multiply performance. 
For the GPU, high aspect ratios in either direction result in significant performance losses, while the IPU is much more stable for such configurations.
The sudden drop in performance on the IPU is probably a compiler issue when using poplin.

\textbf{Observation 2}: Evaluation results show promising performance for the IPU in different scenarios, especially for linear algebra operations based on skewed matrices or sparse matrices.

\subsection{Memory usage for the IPU}

\begin{figure*}[htbp]
	\centerline{\includegraphics[width=0.7\textwidth]{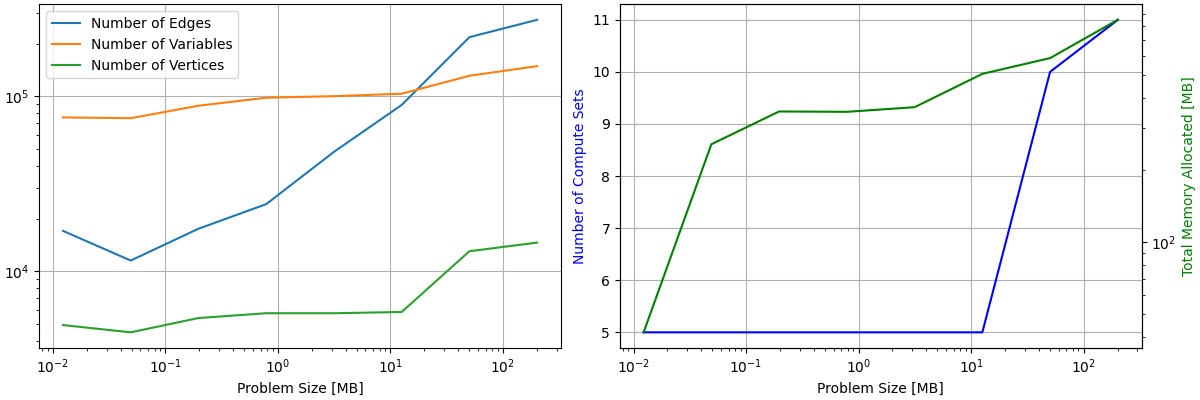}}
	\caption{How different MM problem sizes affect the number of edges, variables, vertices, and available memory on IPU}
	\label{fig: IPU_Memory_Footprint1}
\end{figure*}

A set of experiments is conducted to evaluate how IPU memory requirements change in reaction to an increasing problem size. 
The results are shown in Fig. \ref{fig: IPU_Memory_Footprint1}. 
It is obvious that when working large matrices and problem sizes, there will be exponentially more memory usage.
According to our experiments, it seems that the number of compute sets, which is usually determined by the compiler, substantially impacts memory usage.

\textbf{Observation 3}: The overall memory usage for the IPU does not only depend on the problem size, i.e. the footprint of the corresponding data structures, but there are additional effects with substantially increase overall memory usage. Essentially, this amplifies the need for memory-efficient methods and algorithms.

\subsection{Discussion}

The three preceding subsections leads to three observations.
In summary,
IPUs show a computational performance which is at least on-par with GPUs as long as the task fits into the IPU memory.
For many cases, including sparse or dense skewed linear algebra, the IPU surpasses GPU performance substantially. 
However, memory limitations can severely impact the performance of an IPU.
Additionally, there is a substantial amount of overhead in memory demand, depending on execution parameters such as the number of compute sets which are not under control by the user.

Although TC brings GPU peak performance closer to that of the IPU, it entails structural prerequisites. TC performance degrades faster than GPU performance without TC for skewed matrices, and the vendor does not yet support sparse computations with TC. Moreover, TC's nature imposes greater constraints on required sparsity. Notably, the IPU demonstrated better performance utilization for a given sparsity compared to a GPU.

As neural architectures grow in complexity and size, with a corresponding demand for larger memory capacity, we conclude this section by the finding that IPUs might benefit a lot from techniques that sparsify data structures in neural networks, even if it comes at the cost of more unstructured computations and memory accesses.

\section{Butterfly Factorizations on the IPU}

To assess whether butterfly and pixelfly factorizations can accelerate computations not only on GPUs but also on the GC200 IPU, we conduct evaluations on both platforms. The evaluation involves two primary experiments:
first, we compare the performance of butterfly and pixelfly against a plain \texttt{torch.nn.Linear} computation, for different problem sizes.
Afterwards, we evaluate training time, prediction accuracy and memory compression rate on the respective device for a simple multi-layer perceptron on the CIFAR-10 image classification task.



\subsection{Characterization of Butterfly and Pixelfly Factorizations}
In the following, we examine if factorizations based on butterfly and pixelfly outperform a dense MM (\texttt{torch.nn.Linear}) in terms of processing speed, thereby assessing the amount of overhead for such methods.
In this regard, we measure the execution time of \texttt{torch.nn.Linear}, butterfly and pixelfly with identical matrix dimensions on the A30 GPU and GC200 IPU.

PyTorch implementation was used on both GPU and IPU for both butterfly and pixelfly, since PyTorch can be directly included in PopTorch with minor changes.
Since the time for data movement cannot be directly measured in PopTorch, we run the experiment $1000$ times and take the mean execution time, with identical settings for the GPU experiments. 
We assume that without data movement, the following performance differences would be more drastic. 



\begin{figure*}
	\includegraphics[width=0.8\textwidth]{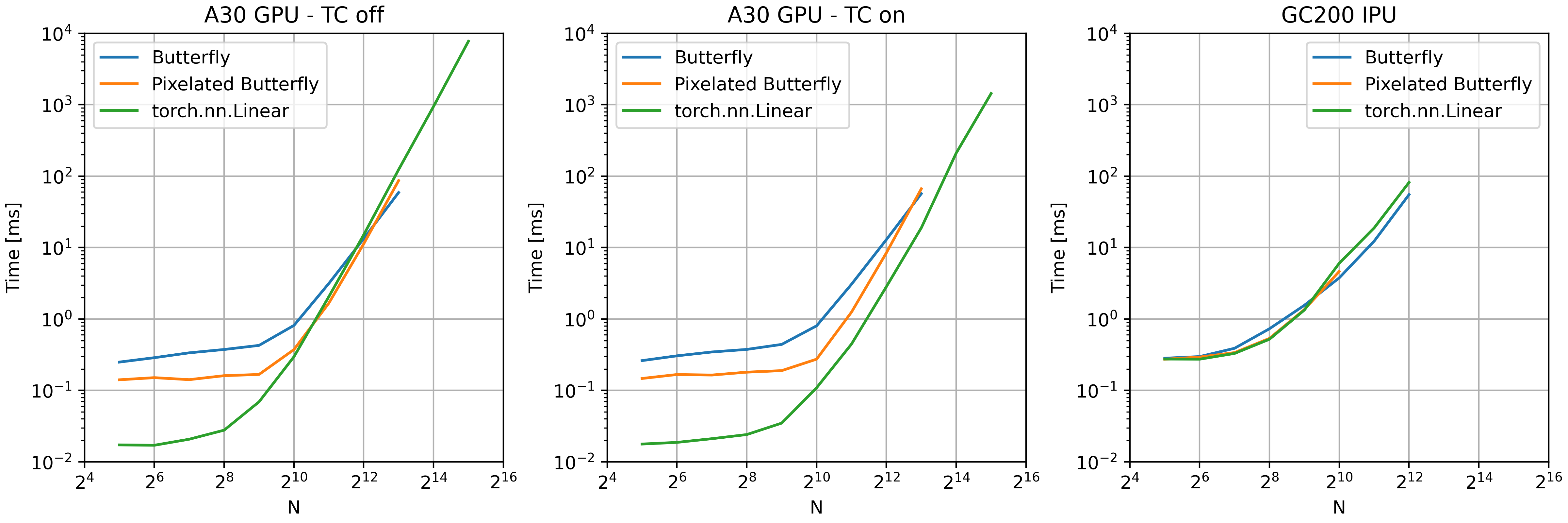}
	\caption{Comparison of \texttt{torch.nn.Linear}, butterfly and pixelfly on the GPU for different matrix dimensions $N$.}
	\label{fig: Speedup_over_LNN_GPU_IPU}
\end{figure*}

Fig.\ref{fig: Speedup_over_LNN_GPU_IPU} shows the execution time for \texttt{torch.nn.Linear}, butterfly and pixelfly on the GPU, both with Tensor Cores turned off (left plot) and on (middle plot) for square dense matrices, as well as the execution time of \texttt{torch.nn.Linear},butterfly and pixelfly in PopTorch on the IPU (right plot).
For the GPU, one can observe that below a certain matrix dimension $N$, \texttt{torch.nn.Linear} outperforms the two factorizations, highlighting that factorization comes with a certain overhead.
In contrast, the experiments on the IPU show much less overhead for the two factorizations in comparison to \texttt{torch.nn.Linear}.
In numbers, the GPU experiments result in a speedup factor of less than $1$ for matrix dimensions $N<2^{11}$, with a worst-case performance degradation of $14.45x$ and $8.8x$, for butterfly and pixelfly, respectively. 
The IPU's break-even point is at $N=2^{10}$, with the worst performance degradation factor being $1.4x$ for butterfly and $1.03x$ for pixelfly.

When dealing with larger problem sizes, the extra computational effort required for the butterfly factorization proves to be beneficial, resulting in improved performance. This observation substantiates the claims made in \cite{butterfly} concerning the advantages of using the IPU. 
As shown in Fig. \ref{fig: Speedup_over_LNN_GPU_IPU}, pixelfly on the GPU exhibits similar performance to \texttt{torch.nn.Linear}, though the latter reaches its limit earlier due to memory limitations. 
We attribute this to the additional memory requirements arising from the low-rank terms and the code necessary for the flat block butterfly.

Compared to the results on the GPU, the behavior of butterfly on the IPU is similar. 
For matrix dimensions greater than $N = 2^{11}$, butterfly outperforms \texttt{torch.nn.Linear}. 
However, in contrast to the GPU results, the benefits of using butterfly is rather negligible. 
The maximum speedup of butterfly and pixelfly is $1.6x$ and $1.3x$, respectively. 
We assume that this can be explained by the Accumulating Matrix Product (AMP) units of the IPU, as these only accelerate \texttt{torch.nn.Linear}, making this hardware unit similar to the Tensor Cores on the GPU. 
Pixelfly performs worse than \texttt{torch.nn.Linear} for every given problem size $N<2^{10}$. 
For matrix dimensions $N<2^{10}$, pixelfly performs better than butterfly. 

\begin{figure}[h]
	\includegraphics[width=\columnwidth]{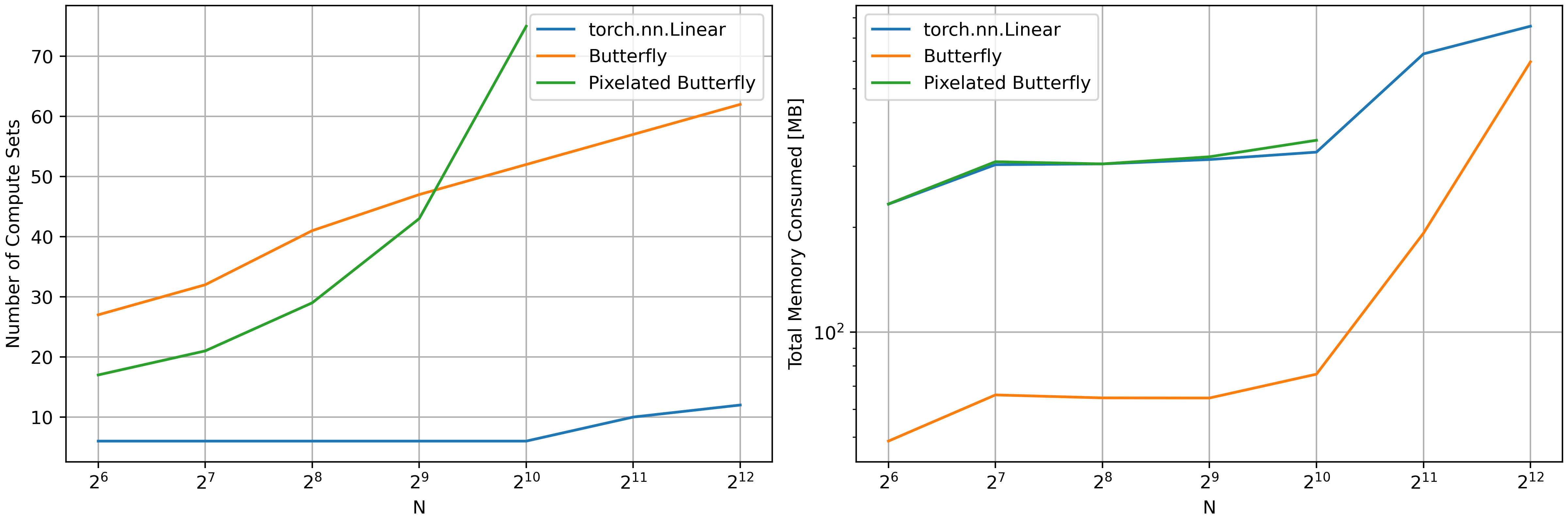}
	\caption{Number of compute sets 
		on the IPU with square matrix dimensions}
	\label{fig:factorization_analysis_IPU}
\end{figure}

To analyze the performance discrepancies between butterfly and pixelfly on the IPU, we utilize the PopVision Graph Analyzer\footnote{Unlike previous experiments, we incorporate a single iteration, due to the overhead of profiling and unpredictable IPU-Fabric errors resulting from slurm integration issues.}.
Fig.\ref{fig:factorization_analysis_IPU} illustrates the relationship between the number of compute sets and the total memory consumption relative to the problem size.
Consistent with earlier findings, the number of compute sets exhibits a significant correlation with the number of variables, edges, and vertices within a specific computation. 

In conclusion, a comprehensive evaluation of IPU performance demands the integration of both computation and memory metrics. 
In the context of the IPU, butterfly factorization emerges as a viable option due to its advantages in terms of reduced memory consumption and enhanced performance. 
On the other hand, when assessing pixelfly, the augmented workload associated with the flat block butterfly and the additional low-rank terms escalates space complexity, rendering this approach less feasible for IPU.

\subsection{CIFAR-10 Image Classification}

In the following, an exemplary real-world ML workload is considered to evaluate the effectiveness and performance of butterfly and pixelfly on both GPU and IPU. 
We follow the methodology of \cite{thomas2018learning} by using a single-hidden layer (SHL) neural architecture on the  CIFAR10 image classification task, and employ a variety of methods for sparsity. 
In more detail, we vary the mini-batch size as well to allow a wider range of comparisons\footnote{In contrast, in \cite{butterfly} the mini-batch size is set to $2^{8}$.}. 
The experiments are conducted by iterating $1000$ times and the average execution time is used to avoid warm-up issues and to stabilize the results. 
In order to measure the execution time of the layers exclusively, we do not measure the overall time for the forward and backwards path. 
Table \ref{table:hyperparams} summarizes the chosen hyperparameters, which in general follow the methodology in \cite{butterfly}. 
Some experiments from \cite{butterfly} could not be incorporated into our experiments due to compatibility issues with the IPU's FFT library in PyTorch.

\begin{table}
	\centering
	\begin{tabular}{l c l c}
		\hline
		Hyperparameter & & Hyperparameter \\
		\hline
		Learning rate & $0.001$ & Activation function & ReLU \\
		Optimizer & SGD & Loss function & Cross-Entropy \\
		Batch size & $50$ & \multirow{2}{*}{Validation set}  &  $15\%$ of   \\
Momentum & $0.9$ &  & training set  \\
		\hline
	\end{tabular}
	\caption{Hyperparameters for SHL benchmark}
	\label{table:hyperparams}
	 \vspace{-4mm}
\end{table}


Table \ref{table:compression_gpu_ipu} shows the results in terms of overall training time and achieved test accuracy for various sparsity methods, as well as a plain training baseline.
For the GPU, results both with and without Tensor Cores are included.
Apparently, the resulting number of parameters $N_{Params}$ is identical for GPU and IPU experiments.

As one can see, butterfly variants show the lowest accuracy loss among the sparsity methods for both datasets, with less than $1.33\%$ loss in accuracy in comparison to the uncompressed baseline for CIFAR-10.
We also conducted a variety of experiments on MNIST, which we omit since insights are mostly inline with those for CIFAR-10.
One notable exception is that for MNIST slight accuracy improvements for butterfly are visible, most likely to improved regularization as a side effect.
While in general, accuracy does not differ a lot for different processor configurations, slight differences ($<1.5\%$) are present and are most likely a result of the non-associativity of floating point beyond instruction boundaries, as well as different weight initializations due to randomization.
In terms of 
$N_{Params}$, it is evident that butterfly variants can reach a high compression ratio compared to the baseline, while maintaining an acceptable accuracy, in particular in direct comparison to similar methods like Fastfood, Circulant, and Low-rank. 

With regard to butterfly and pixelfly on the CIFAR-10 task, one can see that for the GPU pixelfly results in a faster overall training time, resulting in a speed-up of $1.17x$ and $1.10$, respectively without and with using Tensor Cores. 
In contrast, the IPU observes a speed-up of $0.53x$ for the same setting, indicating that the additional computations for pixelfly are resulting in a substantial overhead.
For butterfly,  the compression ratio is 98.5\%, and the training time is $1.62x$ faster on IPU compared to GPU.
However, for pixelfly, the training time on IPU is $1.28x$ slower than on GPU.

This fundamental different behavior is extremely interesting, and it can be attributed to the memory alignment of the block butterfly.
A GPU can be characterized as a dense processor, which can support to some extend block-sparse computations with reasonable performance.
In contrast, the IPU is a sparse processor with much less structural constraints, as shown in the previous section.

While the GPU substantially gains from the benefits of pixelfly with regard to structure, the IPU does not share the same advantage but rather suffers from the computational overhead. 
We conclude this experiment that structured sparsity (pixelfly) comes at additional cost but is of extreme importance for dense processors such as a GPU with and without Tensor Cores, however, a sparse processor such as the IPU suffers from the additional overhead in terms of compute and memory while there is no direct benefit from structural aspects such as memory alignment.

As the pixelfly approach did not work on the MNIST dataset due to the requirements of the matrix sizes being a power of two, we cannot perform a similar comparison for this task.
However, one can see that the general trend of the overhead of butterfly for the two GPU configurations holds true, indicating a similar overall behavior.
In general, as CIFAR-10 is considered a substantially more complex task than MNIST, we did not put additional effort on MNIST.

\begin{table}[h]
	\caption{Single-Hidden-Layer (SHL) benchmark on CIFAR10 dataset with different structured matrix methods compared to baseline matrix approach on GPU and IPU}
	\label{table:compression_gpu_ipu}
	\resizebox{\columnwidth}{!}{%
		\begin{tabular}{lr|rrr|rrr}
			\hline
			\multicolumn{1}{c}{}       & \multicolumn{1}{c|}{}        & \multicolumn{3}{c|}{Accuracy {[}\%{]}}                                                     & \multicolumn{3}{c}{Execution Time {[}s{]}}                                      \\ \cline{3-8} 
			\multicolumn{1}{c}{}       & \multicolumn{1}{c|}{}        & \multicolumn{2}{c}{GPU}                                & \multicolumn{1}{c|}{IPU} & \multicolumn{2}{c}{GPU}                               & \multicolumn{1}{c}{IPU} \\ \cline{3-4} \cline{6-7}
			\multicolumn{1}{c}{Method} & \multicolumn{1}{c|}{N\textsubscript{Params}} & \multicolumn{1}{c}{w/ TC} & \multicolumn{1}{c}{w/o TC} & \multicolumn{1}{c|}{}    & \multicolumn{1}{c}{w/TC} & \multicolumn{1}{c}{w/o TC} & \multicolumn{1}{c}{}    \\ \hline
			\textit{Baseline}          & \textit{1059850}             & \textit{43.94}            & \textit{43.4}              & \textit{44.7}            & \textit{50.43}           & \textit{49.46}             & \textit{24.69}          \\
			Butterfly                  & 16390                        & 42.27                     & 40.75                      & 41.13                    & 61.93                    & 61.46                      & 37.73                   \\
			Fastfood                   & 14346                        & 38.64                     & 37.94                      & 37.68                    & 53.55                    & 51.15                      & 60.70                    \\
			Circulant                  & 12298                        & 28.74                     & 29.21                      & 28.40                     & 54.26                    & 53.92                      & 21.82                   \\
			Low-rank                   & 13322                        & 18.64                     & 18.49                      & 18.59                    & 49.71                    & 53.21                      & 21.75                   \\
			Pixelfly        & 404490                       & 42.61                     & 43.31                      & 43.79                    & 52.79                    & 56.01                      & 71.62                   \\ \hline
		\end{tabular}%
	}
\end{table}


\section{Parameter Sweep for Pixelfly}

Last, we are interested to which extend the performance of pixelfly factorization depends on the chosen parameters for such an approximation.
In particular, we want to make sure that the previously reported performance of pixelfly on the IPU is representative, i.e., that we are not missing particular parameter sets that would improve performance.

We thus briefly study the implications of parameter choices, notably low rank size
on the resulting butterfly size and also the model's overall accuracy.
While a vast amount of experiments have been conducted, we summarize the results by making two of three parameters constant, and only varying the third. 
We do this for every combination of the constant parameters and extract the maximum standard deviation. 
The results are presented in Table  \ref{table:parameter_space_ipu}.

\begin{table}[h]
	\caption{Comparison of mean and standard deviation of metrics when varying parameters on the IPU}
	\label{table:parameter_space_ipu}
	\resizebox{\columnwidth}{!}{%
		\begin{tabular}{ccclrr}
			\hline
			\begin{tabular}[c]{@{}c@{}}Butterfly\\ size\end{tabular}        & \begin{tabular}[c]{@{}c@{}}Block\\ size\end{tabular}            & \begin{tabular}[c]{@{}c@{}}Low-Rank\\ size\end{tabular}         &     \multicolumn{1}{c}{Metric}           & \multicolumn{1}{c}{mean}    & \multicolumn{1}{c}{std}    \\ \hline
			\multirow{3}{*}{var.} & 2\textsuperscript{3}                    &                     & Time[s] & 372    & 107   \\
			& 2\textsuperscript{4}                    & 2\textsuperscript{1}                    & Accuracy[\%]       & 43.8  & 2.2  \\
			& 2\textsuperscript{5}                    &                     & N\textsubscript{Params}        & 1064970 & 326625 \\ \hline
			                    & \multirow{3}{*}{var.} & 2\textsuperscript{2}                    & Time[s] & 465    & 192   \\
			2\textsuperscript{1}                    &                       & 2\textsuperscript{6}                    & Accuracy[\%]       & 38.9  & 1.4  \\
			                    &                       & 2\textsuperscript{7}                    & N\textsubscript{Params}        & 81930   & 184638 \\ \hline
			2\textsuperscript{2}                    & 2\textsuperscript{4}                    & \multirow{3}{*}{var.} & Time[s] & 465    & 18    \\
			2\textsuperscript{7}                    & 2\textsuperscript{3}                    &                       & Accuracy[\%]       & 37.8  & 2.7  \\
			2\textsuperscript{4}                    & 2\textsuperscript{4}                    &                       & N\textsubscript{Params}        & 344074  & 181317 \\ \hline
		\end{tabular}%
	}
\end{table}


Regarding execution time, the influence of the low rank size is relatively minimal, as indicated by $max_{std} = 18$. This outcome aligns with expectations, considering that the low rank term is realized through a dense matrix multiplication. The IPU superior throughput for larger problem sizes contributes to this effect. 
As it also has the highest impact on the test accuracy with a standard deviation of $2.7\%$, it is recommended to set the low rank size to the maximum.
The greatest impact on execution time is created when varying the block size with $max_{std} = 192$. 
The butterfly size has the biggest impact on the number of parameters with $max_{std} = 184638$.
We conclude that for the SHL benchmark with the CIFAR10 dataset, it is beneficial to adapt the configuration of pixelfly depending on the target parameter. 
There is no configuration being optimal with regard to execution time, test accuracy, and parameter count at once, hence, it has to be chosen depending on the primary target. 


\section{Conclusion and Future Works}
GPUs have gained extensive traction across diverse scientific domains, owing to their remarkable capacity for enhancing performance. 
They notably excel in specific tasks, heralding substantial performance enhancements. 
Conversely, the IPU emerges as an innovative accelerator, purposefully crafted to cater to the requirements of ML applications centered on neural networks.

IPUs exhibit a computational power that is comparable to GPUs, provided the computational demands align with IPU memory constraints. 
As neural architectures progressively expand in intricacy and scale, necessitating greater memory capacity, IPUs stand to gain significantly from methodologies aimed at compressing the basic data structures within neural networks. Butterfly and pixelfly are good samples of compression methods used for ML workloads. 

In this paper, a comparison of implementation of butterfly and pixelfly factorization on both GPU and IPU was studied. 
Results are extremely interesting, as they highlight the fact that block-sparse processors such as GPUs require particular sparsity methods that maintain a certain structure in their computations and memory accesses.
In contrast, a sparse processor like the IPU with much less constraints on structure in the sparsity pattern requires different methods. 
In particular, it is not able to exploit any benefits from structure in compute and memory, while it suffers from overhead usually found in methods that gear towards structured sparsity.


Considering that IPU demonstrate excellent performance and support for sparsity, but substantially suffer from memory capacity constraints, we plan to further investigate sparse methods with regard to reduction of state and overhead in compression.
In this regard, we are most interested in scaling to multiply IPUs and the use of streaming memory in combination with sparse methods for scalable learning problems, with both data sets and neural architectures of substantially more complexity than the ones covered here in initial experiments.


\bibliographystyle{ACM-Reference-Format}
\bibliography{ipu_butterfly}

\appendix

\end{document}